\documentclass[12pt]{article}  
  
\def\12{\frac{1}{2}}
\def\14{\frac{1}{4}}

\def\tr{\mbox{tr}}

\def\tr{\mbox{tr}}

\newcommand{\be}{\begin{equation}} 
\newcommand{\ee}{ \end{equation}}
\newcommand{\ba}{\begin{array}}
\newcommand{\ea}{\end{array}}
\newcommand{\bea}{\begin{eqnarray}}
\newcommand{\eea}{\end{eqnarray}}

\def\tr{\mbox{tr}}

\begin{document}  
  
\begin{titlepage}

\begin{flushright}  
 
HUB-EP-01/18\\
   
\end{flushright}  
  
\vspace{3cm}  
\begin{center}  
  
{\Large \bf{Comments on the Energy-Momentum Tensor in 
 Non-Commutative Field Theories}}  
  
\vspace{1cm}  
  
Mohab Abou-Zeid\footnote{abouzeid@physik.hu-berlin.de} and Harald 
Dorn\footnote{dorn@physik.hu-berlin.de} 

\vspace{.3cm}  
{\em Institut f\"{u}r Physik, Humboldt Universit\"{a}t zu Berlin, 
Invalidenstrasse 110, \\ D-10115   Berlin, Germany}   

\vspace{2cm}  
\begin{abstract}
\noindent
In a non-commutative field theory, the energy-momentum tensor obtained from 
the Noether method needs not  be
symmetric; in a massless theory, it needs not be traceless either. In a
non-commutative scalar field theory, the method yields a 
locally conserved yet non-symmetric energy-momentum tensor whose  trace does
not vanish for massless fields. A non-symmetric  tensor also governs  
the response of the 
action  to a general coordinate transformation. In 
non-commutative gauge theory, 
if translations are suitably combined with gauge transformations, the method
yields a covariantly constant tensor which is symmetric but only gauge
covariant. Using suitable Wilson functionals, this can be improved to 
yield a locally conserved and gauge invariant, albeit non-symmetric,
energy-momentum tensor.

\end{abstract}
  
\end{center}  
\vspace{.5cm}
 
\flushleft{April, 2001}
 
\end{titlepage}

\section{Introduction}

Ordinary field theory on a flat 
non-commutative space-time in which the coordinates 
satisfy the commutation relations
\be
[x^\mu ,x^\nu ] =i\theta^{\mu\nu}
\ee
(where $\theta^{\mu\nu}$ is a constant real antisymmetric matrix)
can alternatively be described as field theory on an ordinary space-time but 
with a Moyal deformed (non-commutative) product structure. Many classical,
perturbative and non-perturbative aspects of non-commutative field
theories have been studied (see e.g.~\cite{TF,MRS,RS,GMS,IIKK,MAD} and 
references therein). They also arise in a certain limit of string 
theory (see e.g.~\cite{CDS,SW,AD}). 

In this note we comment on the 
definition and properties of the energy-momentum tensor in such theories. 
This issue has previously been discussed for scalar $\phi ^4$-theory 
in~\cite{AM,PX,GGGPSW}. For D-branes
in a constant background $B$ field in bosonic and superstring theory, the 
symmetric energy-momentum tensors were
constructed in~\cite{OO1,OO2}.

\section{Non-Commutative Scalar Field Theory}

Consider first the non-commutative  real scalar field theory in a flat
four
dimensional space-time of Minkowski signature described by the
action
\be
S= \int dx \left( \frac{1}{2} \partial_\mu \phi \star \partial^\mu \phi
-\frac{m^2}{2}\phi \star \phi -\frac{\lambda}{4!}\phi \star\phi \star\phi
\star \phi \right)~,
\label{scalarS}
\ee
where $\star$ denotes the standard Moyal $\star$-product. In the
following, we will make 
repeated use of the cyclic property of the integration over space-time.
We will also use the
products $\star '$ (introduced in~\cite{LM,MW})  and 
$\star ''$ (introduced in~\cite{PX}) defined by
\be
(f\star ' g) (x) \equiv  \frac{\sin \left( \frac{1}{2}\frac{\partial}
{\partial x_\mu}\theta^{\mu\nu}
\frac{\partial}{\partial y_\nu}\right ) }{\frac{1}{2}\frac{\partial}
{\partial x_\mu}\theta^{\mu\nu}
\frac{\partial}{\partial y_\nu}}~f(x) ~g(y)\vert _{y=x}~ 
\ee
and
\be
(f\star '' g) (x) \equiv  \frac{\cos \left( \frac{1}{2}\frac{\partial}
{\partial x_\mu}\theta^{\mu\nu}
\frac{\partial}{\partial y_\nu}\right ) -1}{\frac{1}{2}\frac{\partial}
{\partial x_\mu}\theta^{\mu\nu}
\frac{\partial}{\partial y_\nu}}~f(x) ~g(y)\vert _{y=x}~.
\ee
All three products $\star$, $\star '$ and $\star ''$  satisfy the Leibnitz
rule. Moreover, $\star$ is associative and non-commutative, $\star '$ is 
non-associative and commutative,  and $\star ''$ is non-associative
and anti-commutative. Furthermore, 
they are related by the formulas
\be
\theta^{\mu\nu} \partial_\mu f\star ' \partial_\nu g = -i
[f,g]_\star ~
\label{2stars}
\ee
and
\be
\theta^{\mu\nu} \partial_\mu f\star '' \partial_\nu g = 
\{ f,g \}_\star -2fg .
\label{rel2}
\ee
Here we defined the Moyal brackets
\be
[ f,g ]_{\star} \equiv  f\star g -g\star f , \ \ \ \ \{ f,g 
\}_{\star} \equiv  f\star g +g\star f .
\ee

The field equation which follows from  action~(\ref{scalarS}) is
\be
\partial^2 \phi +m^2 \phi +\frac{\lambda}{3!} \phi \star\phi\star\phi =0~.
\label{fephi}
\ee
Applying 
the Noether method 
to the action~(\ref{scalarS}) one finds 
the  equation~\cite{AM,PX,GGGPSW}
\be
\partial _\mu T^{\mu\nu}~=~\frac{\lambda}{4!}\left[ [\phi, \partial ^\nu
\phi ]_\star , \phi\star\phi \right]_\star ~,
\label{nocons}
\ee
 where
\be
T_{\mu\nu} = \frac{1}{2} \{ \partial_\mu \phi ,  \partial_\nu \phi \}_\star
-\eta_{\mu\nu} \left( \frac{1}{2}
\partial_\rho \phi \star \partial^\rho \phi -\frac{m^2}{2}\phi\star\phi -
  \frac{\lambda}{4!}\phi \star\phi \star\phi \star \phi \right) .
\ee
If one chooses $T_{\mu\nu}$ as the energy-momentum tensor, then 
energy-momentum is not
conserved locally. On the other hand, due to the fact that
integrals over $\star$-commutators vanish, one still finds global conservation
of energy and momentum \cite{AM,GGGPSW}, at least for theories
without time-space non-commutativity.

Using (\ref{2stars}) twice, together with the antisymmetry of 
$\theta^{\mu\nu}$, the r.h.s. of (\ref{nocons}) can be written as
$-\partial _{\mu}t^{\mu\nu}$ with $t^{\mu\nu}$ defined by
\be
t^{\mu\nu} \equiv \frac{\lambda}{4!}\theta ^{\mu\rho}\theta ^{\alpha\beta}
\left[  a(\partial _{\alpha}\phi\star '\partial _{\beta}\partial ^{\nu}\phi )
\star '\partial _{\rho}(\phi\star\phi )-(1-a)\partial _{\rho}(
\partial _{\alpha}\phi\star '\partial _{\beta}\partial ^{\nu}\phi )\star '(
\phi\star\phi )\right] , 
\ee
where $a$ is a free (real) parameter. This implies that the energy-momentum tensor
\be
{\bf T}^{\mu\nu}~\equiv~T^{\mu\nu}~+~t^{\mu\nu}
\label{TfatT}
\ee
is locally conserved\footnote{The case $a=0$ has been considered in 
\cite{PX}.},
\be
\partial _{\mu}{\bf T}^{\mu\nu}~=~0~.
\label{localcons}
\ee

For $m=0$, the trace of the symmetric part of~(\ref{TfatT}) 
can be cancelled by an
additional term 
which does not contribute to its divergence. The \lq
improved' tensor~\cite{CCJ}
\be
{\bf T}^I_{\mu\nu} \equiv {\bf T}^{m=0}_{\mu\nu}+\frac{1}{6} \left( 
\eta_{\mu\nu}
\partial^2 -\partial_\mu \partial_\nu \right) \phi\star\phi
\ee
has vanishing divergence, however its trace does not vanish; instead
 it is given
by 
\be
t^{\mu}_{\mu} = \frac{\lambda}{4!}\theta^{\mu\rho}\theta ^{\alpha\beta}
\left[ a(\partial _{\alpha}\phi\star '\partial _{\beta}\partial _{\mu}\phi )
\star '\partial _{\rho}(\phi\star\phi )-(1-a)\partial _{\rho}(
\partial _{\alpha}\phi\star '\partial _{\beta}\partial _{\mu}\phi )\star '(
\phi\star\phi )\right] ~.
\ee
This curious  nonvanishing  trace in a massless theory is allowed because the 
theory is not scale invariant (recall that the non-commutativity parameter
$\theta$ has dimensions of length squared)\footnote{By contrast, the authors 
of~\cite{GGGPSW} attribute the
violation of scale invariance to the local non-conservation of the
symmetric tensor $T^{\mu\nu}$.}.

The energy-momentum tensor ${\bf T}_{\mu\nu}$ is locally
conserved, but it is not 
symmetric because of the
 non-symmetric contribution $t_{\mu\nu}$. Although 
this may appear unphysical at first
sight, a closer look shows that the usual arguments for symmetry  do
not apply here. In ordinary field theory, 
the symmetry of
the energy-momentum tensor is a consequence of the requirements that the
angular momentum density is locally conserved and expressible by the familiar 
formula
\be
M^{\mu\nu\rho} = x^\mu {\bf T}^{\nu\rho} - x^\nu {\bf T}^{\mu\rho}~. 
\label{dreh}
\ee
In the present
case, however, the theory is not Lorentz invariant. In particular, 
the non-commutativity 
parameter 
$\theta^{\mu\nu}$ generically breaks the rotational part of the Lorentz
group, so that the angular momentum density is not 
conserved. If~(\ref{dreh}) is imposed, then the energy-momentum tensor
is necessarily non-symmetric. On the other hand, if~(\ref{dreh}) is
abandoned, no conclusion can be drawn concerning
 the symmetry of the energy-momentum
tensor\footnote{In ref.~\cite{PX}, the energy-momentum tensor is
non-symmetric and formula~(\ref{dreh}) is modified.}.

\subsection{The Response to Diffeomorphisms}

Another argument leading to the same conclusion is as follows. Usually, the 
symmetry of the energy-momentum tensor
comes from the way the theory has to couple to gravity.
We do not know how this is  realized in our 
non-commutative field theory, but  
we can investigate the related issue of   
the response of the flat space version to general coordinate 
transformations. Consider the variation
of a Lagrangian density $L[\phi ]$ under an 
infinitesimal coordinate 
transformation $x'=x+\epsilon (x)$. Assuming that $\phi$
is a coordinate scalar, its variation is $\delta\phi =-\cal L_{\epsilon }\phi =
-\epsilon ^{\mu}\partial _{\mu}\phi $ and moreover
\be
L[\phi '(x')]~-~L[\phi (x)]~=~\epsilon ^{\mu }\partial _{\mu }L~+~L[\phi (x)
+\delta\phi (x)]~-~L[\phi (x)]~+~O(\epsilon ^2) .
\ee
For the non-commutative scalar theory with action~(\ref{scalarS}), we assume
for the sake of simplicity that the Moyal product $\star$ always 
refers to the original coordinates $x$ (in other words, $\theta^{\mu\nu}$
is treated as  a coordinate tensor). This leads to
\bea
L[\phi '(x')]~-~L[\phi (x)]&=&-\epsilon ^{\nu }\partial ^{\mu } T_{\mu\nu }
+\frac{\lambda}{4!}[[\phi ,
\epsilon ^{\nu}\partial _{\nu}\phi]_{\star},\phi\star
\phi ]_{\star} -\frac{1}{2}\{\frac{\delta S}
{\delta \phi },\epsilon ^{\nu} \partial _{\nu}\phi\}_\star
\nonumber\\
&& +\frac{1}{2}\epsilon ^{\nu }\partial ^{\mu}\{ \partial _{\mu}\phi ,
\partial _{\nu }\phi \} _\star -\frac{1}{2}\partial ^{\mu}\{\partial _{\mu}
\phi ,
\epsilon ^{\nu}\partial _{\nu}\phi\}_{\star}~+O(\epsilon ^2).
\label{response}
\eea
For constant $\epsilon $, due to translational invariance, the l.h.s.
is zero, the fourth and fifth terms on the r.h.s. cancel and $\epsilon ^{\nu}$
in the second  and  third terms can be taken out of the 
$\star$-commutator. On shell, this 
implies eq.~(\ref{nocons}) and its reformulation~(\ref{localcons}), 
respectively. Off shell, this implies the identity
\be
-\partial ^{\mu } T_{\mu\nu }
+\frac{\lambda}{4!}[[\phi ,
\partial _{\nu}\phi]_{\star},\phi\star
\phi ]_{\star} - \frac{1}{2}\{\frac{\delta S}
{\delta \phi }, \partial _{\nu}\phi\}_\star =0.
\label{off}
\ee
On the other hand, we may consider the response of the
action under a  general coordinate transformation. Using~(\ref{off}),  
one finds up to $O(\epsilon ^2)$:
\bea
L[\phi '(x')]~-~L[\phi (x)]&=&\frac{1}{2}\epsilon ^{\nu }
\partial ^{\mu}\{ \partial _{\mu}\phi ,
\partial _{\nu }\phi \} _\star 
-\frac{1}{2}\partial ^{\mu}\{\partial _{\mu}\phi ,\epsilon ^{\nu}
\partial _{\nu}\phi\}_{\star}\nonumber\\
&&+\frac{\lambda}{4!}
\Big ([[\phi ,\epsilon ^{\nu}\partial _{\nu}\phi]_{\star},
\phi\star\phi ]_{\star}-\epsilon ^{\nu}[[\phi ,\partial _{\nu}\phi]_{\star},
\phi\star\phi ]_{\star}\Big ) \nonumber \\ & & -\frac{1}{2}
\left( \{\frac{\delta S}
{\delta \phi },\epsilon ^{\nu} \partial _{\nu}\phi\}_\star -\epsilon ^{\nu}
\{\frac{\delta S}
{\delta \phi }, \partial _{\nu}\phi\}_\star \right) .
\eea
Note that under 
the space-time integral, total derivatives and  Moyal commutators 
will not contribute; moreover, the Moyal product in the penultimate term
can be replaced with ordinary multiplication. Utilizing~(\ref{rel2}), 
we find 
\be
S[\phi '(x')]~-~S[\phi (x)]~=~-\int dx \left( {\bf T}^{\mu\nu} +
\frac{1}{2} \theta^{\mu\rho} \frac{\delta S}
{\delta \phi } \star '' \partial_\rho \partial^\nu \phi \right) 
~\partial_{\mu}\epsilon_{\nu}~+~O(\epsilon ^2)~.  
\label{response}
\ee
This result is valid for arbitrary field configurations and arbitrary 
diffeomorphisms $\epsilon$. Thus the
 response to general coordinate transformations is governed by the 
non-symmetric
but locally conserved  tensor ${\bf T}^{\mu\nu}+ \frac{1}{2} \theta^{\mu\rho} 
\frac{\delta S}
{\delta \phi } \star '' \partial_\rho \partial^\nu \phi$. Although the non-symmetry
is consistent with our previous argument based on broken Lorentz invariance, 
 it is not clear to us whether this result
has any bearing on the existence of a diffeomorphism-invariant 
generalization of the theory~(\ref{scalarS}).

\subsection{The Belinfante Problem}

Although we have just argued that the usual physical reasons for the symmetry
of the energy-momentum tensor do not apply, one nevertheless
may wonder whether a  further contribution can be added to~(\ref{TfatT})
in order to make it symmetric, along the lines of the Belinfante 
procedure~\cite{CJ}.  Consider adding to ${\bf T}_{\mu\nu}$ a contribution of 
the 
form $\partial^\rho \chi_{\rho\mu\nu}$, where the tensor
$\chi_{\rho\mu\nu}$ is antisymmetric in its first two indices,
\be
\chi_{\rho\mu\nu} = \chi_{[\rho\mu ]\nu} .
\ee
Such a contribution would clearly not change the divergence of $T_{\mu\nu}$.
Thus the tensor
\be
{\bf \hat T}_{\mu\nu} \equiv {\bf T}_{\mu\nu} +\partial^\rho \chi_{\rho\mu\nu}
\ee
would also be locally conserved, and it would be symmetric provided  
$\chi_{\rho\mu\nu}$
is a solution to the differential equation
\bea
2 \partial^{\rho} \chi_{\rho}^{[\mu \nu ]} & = & -~\frac{\lambda}{4!} 
\left[ \theta^{\mu\rho}\theta ^{\alpha\beta}
\left( a(\partial _{\alpha}\phi\star '\partial _{\beta}\partial ^{\nu}\phi )
\star '\partial _{\rho}(\phi\star\phi ) \right. \right. \nonumber\\
&&~~~~~ \left. \left. -(1-a)\partial _{\rho}(
\partial _{\alpha}\phi\star '\partial _{\beta}\partial ^{\nu}\phi )\star '(
\phi\star\phi ))~-~(\mu \leftrightarrow\nu  \right) \right].
\label{cond}
\eea
In the case $m=0$, the 
additional condition of vanishing  trace would require that
$\chi_{\rho} \equiv \eta^{\mu\nu}\chi_{\rho\mu\nu}$ satisfies 
\be
 \partial^{\rho} \chi_{\rho}~ =~ -t^{\mu}_{\mu}~.
\ee
There is no violated integrability condition for these two differential
equations, which certainly admit integral solutions. The 
problem consists in finding a solution which
depends on $x$ $only$ via the dependence on the field 
$\phi $ and its derivatives.
There is no solution within an ansatz taking into account
the lowest orders in the number of derivatives and the relevant $\star $ 
and $\star '$ powers of $\phi $. Since there are mixings between different
orders, we cannot exclude the existence of suitable solutions, but
on the basis  of a cursory search we  suspect the absence of any 
solution
of the required type.

\section{Non-Commutative Gauge Theory}

We now turn to the case of non-commutative $U(N)$ Yang-Mills theory in $D$
dimensions, with action
\be
S= \int dx~\tr F_{\mu\nu}\star F^{\mu\nu}~, 
\label{Sgauge}
\ee
where
$F_{\mu\nu}$ denotes the 
strength of the non-commutative Yang-Mills gauge field $A_\mu$. 
In standard Yang-Mills theory the canonical energy-momentum tensor
is neither symmetric nor gauge invariant. By applying the Belinfante 
procedure, one finds a symmetric and gauge invariant improved tensor.
However, there is a more direct way to get the improved tensor by combining 
in the
Noether procedure translations with suitably adapted gauge transformations
\cite{JM}. Applying this method to non-commutative Yang-Mills theory
we get
\be
D_{\mu}~T^{\mu\nu}~\equiv~\partial _{\mu}T^{\mu\nu}~-~iA_{\mu}\star T^{\mu\nu}
~+~iT^{\mu\nu}\star A_{\mu}~=0~, 
\label{YMT1}
\ee
with
\be
T^{\mu\nu}~=~2\{F^{\mu\rho},F^{\nu}_{\ \rho}\}_\star -\eta ^{\mu\nu}
F^{\alpha\beta}\star F_{\alpha\beta}~.
\label{gaugesymmT}
\ee
$T^{\mu\nu}$ is symmetric and expressed in terms of field strength
only. But due to the peculiarities of gauge theory on non-commutative
spaces, it is not gauge invariant. Under a gauge transformation
$\delta A_\mu = D_\mu \Omega$, it transforms  covariantly as
\be
T^{\mu\nu}~\rightarrow ~\Omega\star T^{\mu\nu}\star\Omega ^{-1}~.
\ee
Eq. (\ref{YMT1}) implies that $\partial _{\mu}\tr T^{\mu\nu}$ is equal
to a $\star $ commutator, which gives no contribution after integration
over space-time. In addition, the integrated $\tr T^{\mu\nu}$ is gauge
invariant. Therefore one could use the integral over $\tr T^{\mu\nu}$
to define global gauge invariant conserved 
quantities\footnote{To be more precise, this construction would require
the absence of time-space non-commutativity.}.

Let us be more ambitious and look for a locally conserved and gauge
invariant energy-momentum tensor. For all local operators transforming
covariantly under gauge transformations, one 
can construct gauge invariant Fourier modes
by attaching a suitably  adapted Wilson functional \cite{DR,GHI}.
Transforming back to ordinary space one then arrives at local
gauge invariant quantities \cite{BL}. For our $T^{\mu\nu}$, 
this yields the gauge
invariant tensor
\be
\hat T^{\mu\nu}(y)~=~\frac{1}{(2\pi )^D}\int dkdx~e^{iky}
e^{-ikx}\star \tr\big ( U(k,x)\star T^{\mu\nu}(x)\big )~,
\label{hatT}
\ee
with 
\be
U(k,x)~=~P_{\star}\exp \left (i\int _0^1 dt A_{\mu}(x+t\theta k)
\theta^{\mu\nu} k_{\nu}\right )~.
\ee 
$P_{\star}$ denotes path ordered $\star$ multiplication,
from right to left with increasing contour parameter. The $\star$
multiplication refers to the functional dependence on $x$.

As a side remark, let us mention that this construction becomes 
more natural if one takes  into account the remarkable 
identity~\cite{DW,BLP,BL}
\be
e_{\star}^{-ik\hat x}~=~e^{-ikx}\star U(k,x)~.
\label{estar}
\ee
The subscript $\star $ on the exponential on the left hand side
indicates that the multiplication in the  Taylor series of its argument is
$\star $~. Here $\hat x^{\mu} =x^{\mu}+\theta ^{\mu\nu}A_{\nu}$ is the gauge 
covariant coordinate introduced in ref.
\cite{MSSW}. A straightforward
proof of this identity can be given by showing that, as a function
of the scale of $k$, 
$e^{ikx}\star e_{\star}^{-ik\hat x}$ and $U(k,x)$ fulfil the same first 
order differential
equation with the same initial condition. 
Using (\ref{estar}), the construction in (\ref{hatT}) simply corresponds
to going forward and backward into momentum space while replacing
in the first step $e^{-ikx}$ by $e_{\star}^{-ik\hat x}$.

Taking the derivative of (\ref{hatT}) with respect to its argument $y$
and treating the arising factor $k$ under the integral as a derivative
with respect to $x$ yields
\be
\partial _{\mu}\hat T^{\mu\nu}(y)=\frac{1}{(2\pi )^D}\int dkdx~e^{iky}
e^{-ikx}\star \tr \left ( 
\frac{\partial}{\partial x^{\mu}}U(k,x)\star T^{\mu\nu}
+U\star \partial _{\mu}T^{\mu\nu}(x)\right ).
\label{div1}
\ee
The derivative of $U(k,x)$ with respect to $x$ is related to a shift
of the straight line from $x$ to $x+\theta k$ as a whole. 
One finds
\be
\frac{\partial}{\partial x^{\mu}}U(k,x)=iA_{\mu}(x+\theta k)\star U-iU
\star A_{\mu}(x)+iP_{\star}\left (\int _0^1ds F_{\mu\alpha}(x+s\theta k)
\theta ^{\alpha\beta}k_{\beta}U\right ).
\ee
Inserting this into (\ref{div1}) and substituting (\ref{YMT1}), we arrive at
\be
\partial _{\mu}\hat T^{\mu\nu}(y)=\frac{i}{(2\pi )^D}\int dkdx~e^{iky}
e^{-ikx}\star \tr P_{\star}\left ( \int _0^1dsF_{\mu\alpha}(x+s\theta k)\theta
^{\alpha\beta}k_{\beta}U(k,x)T^{\mu\nu}(x)\right ).
\ee
Now the factor $k_{\beta}$ under the integral on the r.h.s. can be thought of
as being generated by a differentiation with respect to $y^{\beta}$. 
This finally  gives us 
the local conservation law
\be
\partial _{\mu}{\bf T}^{\mu\nu}~=~0 ,
\ee
where
\bea
{\bf T}^{\mu\nu}(y)&= &\frac{1}{(2\pi )^D}\int dkdx~e^{iky}
e^{-ikx}\star \tr\Big [ U(k,x)\star T^{\mu\nu}(x)\nonumber\\
&&-~\theta ^{\mu\alpha}P_{\star}\left ( 
\int _0^1ds F_{\alpha\beta}(x+s\theta k)U(k,x)T^{\beta\nu}(x)\right )\Big ] .
\label{final}
\eea 

As in the scalar case, our locally conserved ${\bf T}^{\mu\nu}$
is not symmetric. In four dimensions, $T$ as defined in (\ref{gaugesymmT})
is traceless but ${\bf T}$ has nonvanishing trace.

\section{Summary and Outlook}

To summarize, the application of the Noether method to non-commutative
field theories which admit continuous symmetries yields locally conserved currents. In
particular, invariance of the action under space-time translations yields a locally
conserved energy-momentum tensor. For the non-commutative scalar field theory with
action~(\ref{scalarS}), this is given by eq.~(\ref{TfatT}). The fact that this tensor
is not symmetric does not lead to any obvious contradictions
(at least at the classical level) because the theory is not Lorentz invariant and
the angular momentum density is not conserved. The fact that it is not traceless
for massless fields is likewise allowed because the theory is not scale invariant. If
one considers the variation of~(\ref{scalarS}) under arbitrary infinitesimal
diffeomorphisms under
which $\theta^{\mu\nu}$ transforms like a contravariant antisymmetric tensor, then
the response is governed by a non-symmetric but locally conserved tensor (see 
eq.~(\ref{response})).

In the case of non-commutative gauge theory with gauge group $U(N)$, invariance of 
action~(\ref{Sgauge}) under infinitesimal
translations and gauge transformations yields the  symmetric and covariantly conserved
energy-momentum tensor~(\ref{gaugesymmT}) which however is not gauge invariant. In 
order to construct a gauge invariant and locally conserved energy-momentum tensor,
one may attach Wilson functionals in a suitable manner; this procedure yields
the non-symmetric tensor~(\ref{final}). Finally, we note that, as emphasized by the 
authors of~\cite{OO1},
 their string theoretic construction of an (symmetric and locally conserved) 
energy-momentum tensor and a pure 
field theoretic study are not related in an obvious manner, since the 
resulting tensors  
correspond to different orders in an expansion in the string 
tension. Nevertheless, it 
is remarkable that in our field theoretical analysis
field strength insertions into Wilson functionals also
play a crucial role.   
\\[5mm]
{\bf Acknowledgements}\\
We would like to thank J. Gracia-Bondia for useful discussions and for
pointing 
out 
ref.~\cite{JM}. M.\ A.\ is supported by the Swiss 
National Science Foundation 
under grant 
number 83EU-056178. H.\ D.\ is partially supported by the 
Deutsche Forschungsgemeinschaft.

\end{document}